# Comparison of different thermostats in the Holstein model

*N. Fialko[1], M. Olshevets, V.D. Lakhno*

Institute of Mathematical Problems of Biology RAS – the Branch of Keldysh Institute of Applied Mathematics of Russian Academy of Sciences, Pushchino, Russia

When modeling charge dynamics in a chain of $N$ sites at a temperature $T$, a Langevin thermostat and a Hamiltonian system, i.e., a chain heated to a given temperature before charge is injected, are compared. It is shown that the polaron disruption occurs in the same range of values of the thermal energy $NT$, however, $T$ is not given by the initial data, but obtained after simulation from the average kinetic energy. For large $T$, the results averaged over a set of trajectories in a system with a Langevin thermostat and the results averaged over time for a Hamiltonian system are close, which does not contradict the Ergodic hypothesis.

## INTRODUCTION

Mathematical modeling and computational experiments are important tools for studying charge transfer processes in biopolymers such as DNA. The relevance of research of charge transfer processes in DNA is associated, in particular, with progress in nanobioelectronics, which is a potential substitution for modern microelectronics based on semiconductor technology [1 – 3 and ref. therein].

From a mathematical point of view, modeling of charge transfer in quasi-one-dimensional biomolecules is reduced to the following. The biopolymer is modeled by a chain of sites (groups of strongly bonded atoms), whose motion is described by classical equations of motion. The dynamics of a charge is described by Schroedinger equation. Site displacements affect the probabilities of finding a charge, and vice versa – charge localization at a site leads to its displacement from the equilibrium state.

An important part of modeling is taking into account the influence of the environment. When describing the site displacements, various methods of setting the temperature can be applied (temperature is understood as the average kinetic energy of sites). For the Holstein model, we compared two methods: Langevin thermostat **LT** (when terms with friction and a random force with a special distribution are added to the classical equations of the system, as in refs. [4–7]) and Hamiltonian system **HS**, in which the temperature is specified only by the initial distribution of site velocities and displacements (this method is used, e.g. in [8,9]).

For molecular dynamics, when motion of the particles is described by a system of classical equations, different thermostats have been studied for a long time [10,11]. For the semiclassical model, comparisons of different thermostats, as far as we know, have not been previously carried out.

---

[1] E-mail: fialka@impb.ru

## MODEL

The model is based on Holstein Hamiltonian for a discrete chain of sites in a semiclassical approximation. For a chain of $N$ homogeneous sites the equations of motion in dimensionless form are:

$$i\dot{b}_n = \eta(b_{n-1} + b_{n+1}) + \chi u_n b_n, \quad (1)$$

$$\ddot{u}_n = -\omega^2 u_n - \chi |b_n|^2 \quad (n = 1,...,N) \quad (2)$$

where $b_n$ is the amplitude of the probability of the charge (electron or hole) occurrence at the $n$-th site, and $u_n$ is site displacement from its equilibrium position, $v_n = \dot{u}_n$ is site velocity. The total energy is

$$E_{tot} = E_{kin} + E_{pot} + E_\eta + E_\chi = \frac{1}{2}\sum_n v_n^2 + \frac{\omega^2}{2}\sum_n u_n^2 + \eta\sum_n (b_{n+1}b_n^* + b_n b_{n+1}^*) + \chi\sum_n u_n b_n b_n^*. \quad (3)$$

Here, parameter $\eta$ is matrix elements of the transition between the $n$-th and $(n+1)$-th sites (depending on overlapping integrals), $\omega$ is oscillation frequency of classical sites and $\chi$ is the constant of coupling between quantum (1) and classical (2) subsystems (for more detail of dimensionless model see [7]).

LT was previously considered [6,7], in which subsystem (2) involves the term with friction $\gamma$ and White noise $Z_n(\tilde{t})$, so the equations $\ddot{u}_n = -\omega^2 u_n - \chi|b_n|^2 - \gamma\dot{u}_n + \xi Z_n(t)$ are considered instead of (2) ($\xi^2 = 2 E^* T \gamma$ with undimensional coefficient $E^*$). For the system with LT the averaged "over ensemble" functions of time $\langle E_{tot}(t)\rangle$, $\langle E_{kin}(t)\rangle$, etc were calculated [6,7]. Here we present the results obtained for the HS (1,2), in which the temperature is given only by the initial distribution of site displacements and velocities $\{u_n(0), v_n(0)\}$.

## NUMERICAL SIMULATIONS

The calculations were carried out for chains of length $N = 40$ и $N = 80$ sites. Parameter values $\eta = -0.456$, $\omega = 0.5$, $\chi = 1$ are as for LT [6]. For each $N$ at fixed $T$ (in the range from 0 to 300 K) three sets of initial data $\{u_n(0), v_n(0)\}$ were generated; for each set, different initial charge distributions in the chain $\{b_n(0)\}$ are given: {1} polaron, {2} the charge appearance at one site in the center of the chain $b_{N/2}(0) = 1$, and two variants of uniform distribution of probabilities along the chain $b_n(0) = 1/\sqrt{N}$ {3} and $b_n(0) = (-1)^n/\sqrt{N}$ {4} (here, although the probabilities at $t=0$ $|b_n(0)|^2 = 1/N$ are the same, the energy of the system $E_{tot}$ (3) differs by $\approx 4|\eta|$).

HS (1,2) was numerically integrated by the classical 4th order Runge–Kutta method. The trajectories were integrated over long time intervals, and then the average energies (3) were calculated, including the average kinetic energy

$$\bar{E}_{kin} = \frac{1}{\Delta t}\int_{\Delta t} E_{kin} dt, \quad (4)$$

which was used when the temperature of chain was estimated at the end of the calculation: $T_\infty = 2\bar{E}_{kin}/(NE^*)$ (for the chosen parameters and $T$[K] the scaling factor is the following $E^* \approx 0.001309$ [7]).

Figure 1 shows the time dependencies of energies (3) for initial polaron {1} in a chain $N = 40$ at the initial temperature $T_0 \approx 40$. The graphs demonstrate that the system begins to oscillate around a new equilibrium state, and in this case the kinetic energy decreases, i.e. the chain cools down.

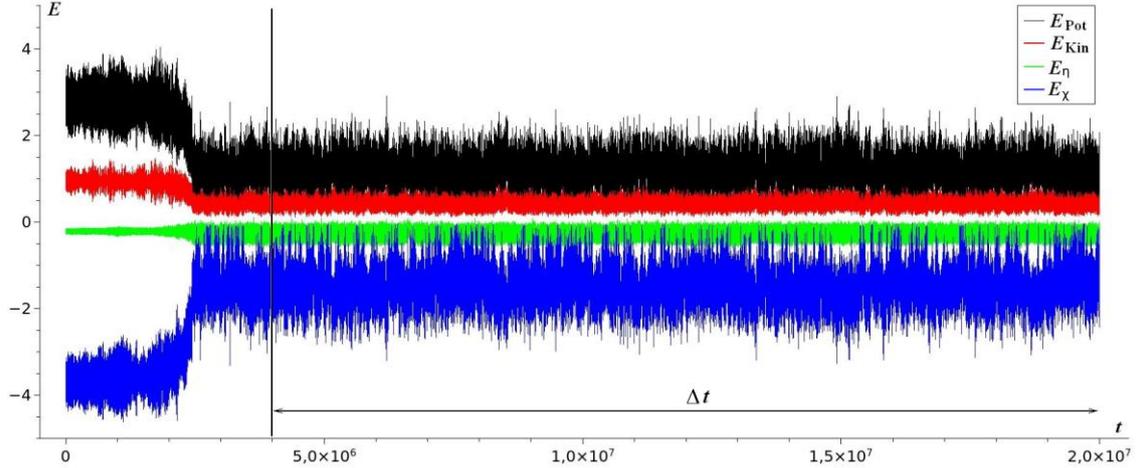

Fig. 1. Time dependencies of energies (3) for polaron {1} at $T_0 \approx 40$, new average value of kinetic energy (4) is calculated over $\Delta t$ (shown by a horizontal arrow) and corresponds to $T_\infty \approx 17$.

According to the simulation results, the transition from the polaron regime to the delocalized state occurs in the same region of the thermal energy of the chain for both LT and HS, but for HS the kinetic energy is not given by the initial data, it is determined after the calculation (4). Figure 2 shows the results of calculations of the electronic part of the energy, which is calculated as the total energy (averaged by realizations for LT, results from [6]) minus the thermal energy of the $N$-site chain without charge (site oscillator satisfies the Equipartition theorem $E_{kin} = E_{pot} = E^*T/2$); for HS thermal energy is estimated from the average kinetic energy (4).

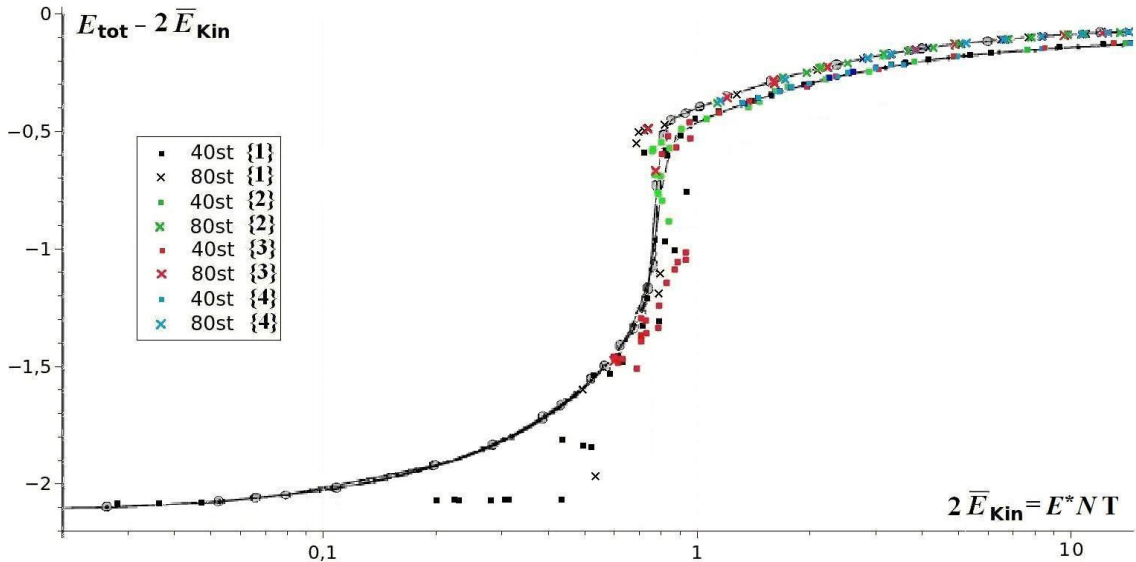

Fig. 2. Dependencies of the electronic part of the energy on the thermal energy of the chain for HS (symbols in different colors indicate different initial data for subsystem (1), squares – for chains of $N = 40$ sites, crosses for $N = 80$) and for LT (line for $N = 40$ sites, line with circles for $N = 80$ [6]).

Figure 2 shows that the transition from the lower branch (polaron) to the upper (delocalized state of charge) for LT and HS occurs in the same region of thermal energy $2\bar{E}_{\text{Kin}} = E^*NT \approx 0.8$. There is quantitative agreement for the upper branch of the graph. For the lower branch (for low temperatures), the difference in the average values for different ways of setting the temperature is significant.

## CONCLUSIONS

Numerical simulations show that for high temperatures the results averaged over a set of trajectories in the LT system and the results averaged over time for the HS are close, which does not contradict the Ergodicity hypothesis. Practically, for homogeneous DNA fragments at biologically significant temperatures $T \approx 300$ K, one can use either of two variants for setting the thermostat.

We are grateful to the Keldysh Institute of Applied Mathematics of the Russian Academy of Sciences for providing high-performance computational facilities of k-60 and k-100 (https://ckp.kiam.ru).